\documentclass[12pt]{iopart}

\usepackage{graphicx}
\usepackage{epsfig}
\usepackage{subfig}
\usepackage{amssymb}
\newcommand{\beginsupplement}{%
        \setcounter{table}{0}
        \renewcommand{\thetable}{S\arabic{table}}%
        \setcounter{figure}{0}
        \renewcommand{\thefigure}{S\arabic{figure}}%
     }
\begin{document}

\title[Network community detection using modularity density measures]{Network community detection using modularity density measures}

\author{Tianlong Chen\textsuperscript{1,2}, Pramesh Singh\textsuperscript{1,2}, Kevin E. Bassler\textsuperscript{1,2,3}}
\address{\textsuperscript{1}~Department of Physics, University of Houston, Houston, Texas 77204, USA.\\
\textsuperscript{2}~Texas Center for Superconductivity, University of Houston, Houston 77204, Texas, USA.\\
\textsuperscript{3}~Department of Mathematics, University of Houston, Houston, Texas 77204, USA.}
\ead{bassler@uh.edu}


\begin{abstract}
Modularity, since its introduction, has remained one of the most widely used metrics to assess the quality of community structure in a complex network. However the resolution limit problem  associated with modularity limits its applicability to networks with community sizes smaller than a certain scale. 
In the past various attempts have been made to solve this problem. More recently a new metric, {\it modularity density}, was introduced for the quality of community structure in networks in order to solve some of the known problems with {modularity}, 
particularly the resolution limit problem. Modularity density resolves some communities which are otherwise undetectable using modularity. However, we find that it does not solve the resolution limit problem completely by investigating some cases where it fails 
to detect expected community structures. 
To address this problem, we introduce a variant of this metric and show that it further reduces the resolution limit problem, effectively eliminating the problem in a wide range of 
networks.

\end{abstract}

\maketitle

\section{Introduction}
An important problem in study of complex graphs is that of characterizing and detecting community structure within them~\cite{newman2004epj,danon2005,fortunato2010}. Processes occurring on networks often depend on the network topology and in particular on the community structure~\cite{singh2013}. Therefore identifying the community structure is essential in understanding and modeling complex systems~\cite{mentzen2008}. Various definitions of community exist~\cite{schaub2017}, with the community structure depending on the definition, and no definition is guaranteed to be the best for all applications~\cite{peel2017}.
Often, however, communities are thought of as groups of nodes that are more densely connected together than they are with nodes in other groups. One of the most widely used metrics to quantify community structure based on this idea is {\it modularity}~\cite{newman2002,newman2003,newman2004}.
For a given partition of the nodes of a network $C = \{c\}$, modularity $Q$ is defined as the fraction of links within communities minus the expected fraction in a corresponding random network that serves as a null model,
\begin{eqnarray}
Q = \sum_{c \in C}\left[\frac{m_c}{m} -\left(\frac{2 m_c + e_c}{2m} \right)^2\right]
\label{Q}
\end{eqnarray}
where $m_c$ is the number of links in community $c$,
$e_c$ is the number of external links of $c$,
and $m$ is the total number of links in the network.
The partition that maximizes $Q$ is considered as the one that corresponds to the community structure.
Community structures based on maximizing $Q$ have been found in a wide variety of networks such as communication, infrastructural, biological, and social networks~\cite{newman2002, newman2004, newman2006pre}.  

Despite its popularity, the metric $Q$ has drawbacks. Perhaps the most notable is that by maximizing $Q$ one may not detect communities that contain fewer links than 
\begin{eqnarray*}
m_c \sim \sqrt{2m}
\end{eqnarray*}
 This is known as the {\it resolution limit} (RL) problem~\cite{fortunato2007}.  A number of approaches have been taken toward  solving this problem~\cite{ronhovde2010,arenas2008,granell2012,aldecova2011}. One approach has been to modulate the relative weights of the two terms in Eq.~\ref{Q}~\cite{arenas2008}. Indeed this approach does allow smaller communities to be detected, but at the cost of then not being able to detect large communities~\cite{lancichinetti2011}. Another approach has been to use a different null model for the second term in Eq.~\ref{Q}~\cite{traag2011}. Doing that though may affect the character of the community structure that will be detected~\cite{lancichinetti2011,fortunato2016}. Perhaps the most promising approach, however, is to use a new metric called {\it modularity density} to quantify community structure.  This measure was recently introduced by Chen et al.~\cite{mingming2013} to address
multiple issues with modularity, particularly the resolution limit problem.
Modularity density $Q_{ds}$ is defined as:
\begin{eqnarray}
Q_{ds} = \sum_{c \in C}\left[\frac{m_c}{m} p_c -\left(\frac{2 m_c + e_c}{2m} p_c \right)^2 - \sum_{ c' \ne c} \frac{m_{cc'}}{2m}p_{cc'}\right]
\label{Qds}
\end{eqnarray}
where
$m_{cc'}$ is the number of links between communities $c$ and $c'$,
$n_c$ is the number of nodes in $c$,
$p_c = 2m_c/[n_c (n_c - 1)]$ is the density of links inside $c$,
$p_{c,c'} = m_{cc'}/(n_cn_{c'})$ is the density of links between $c$ and $c'$,
and the other quantities are the same as in Eq.~\ref{Q}.
Again, it is the partition that maximizes $Q_{ds}$ that corresponds to the community structure.

As can be seen from Eqs.~\ref{Q} and \ref{Qds}, there are two main differences between modularity density and modularity.
The most significant difference is that modularity density adds coefficients related to link densities to each term in its definition. It is for this reason the metric is called what it is. If only the first two terms are considered, then it has been found that a structure consisting of many small communities is often found or resolved, even in situations where using $Q$ fails to do so.
Thus, the RL problem is at least partially mitigated using $Q_{ds}$, but perhaps by creating communities that are too small.
The third term in the definition of $Q_{ds}$, which is its second main difference with modularity, was introduced to help alleviate a tendency for $Q$ and $Q_{ds}$ to find communities that are too small.
This term is referred to as the \textit{Split Penalty} (SP). 

In this paper we show that, although using modularity density does alleviate the RL problem in many cases, it does not completely eliminate it. There is still a RL problem when using $Q_{ds}$.
We show this by identifying limitations of applying $Q_{ds}$ to certain example cases. 
We also show that the SP term can have undesired consequences.
To address these problems discovered with using $Q_{ds}$, 
we propose a new metric to quantify community structure, a variant of modularity density, which we refer to as {\it excess modularity density} $Q_x$. We show that using $Q_x$ further mitigates the RL problem, resolving communities in cases when using either $Q$ or $Q_{ds}$ fails to do so. Also, $Q_x$ has no SP term.

The rest of the paper is organized as follows. In the next section we use $Q_{ds}$ to find the community structure in the Zachary's Karate Club network and discuss potential problems that arise due to the SP term in $Q_{ds}$. In Sec.~3 we use simple networks structures to demonstrate that $Q_{ds}$ also suffers from RL problems. For some specific examples, we identify the conditions under which $Q_{ds}$ becomes unreliable. In Sec.~4, we propose a modified metric $Q_x$ in an attempt to fix the issues with $Q_{ds}$. We test the use of this new metric on a number of networks and observe that $Q_x$ indeed addresses the issues found with $Q_{ds}$. We also discuss the fundamental limitations that even $Q_x$ has with respect to the RL problem.
In the final section, we conclude by summarizing our results, arguing for the superiority of using the density metric $Q_x$, and discussing possible future research directions to further extend the idea and applications of modularity density measures.

\section{Modularity density applied to the Karate Club network\label{SP}}
\begin{figure}
\subfloat[]{\includegraphics[width=0.5\textwidth]{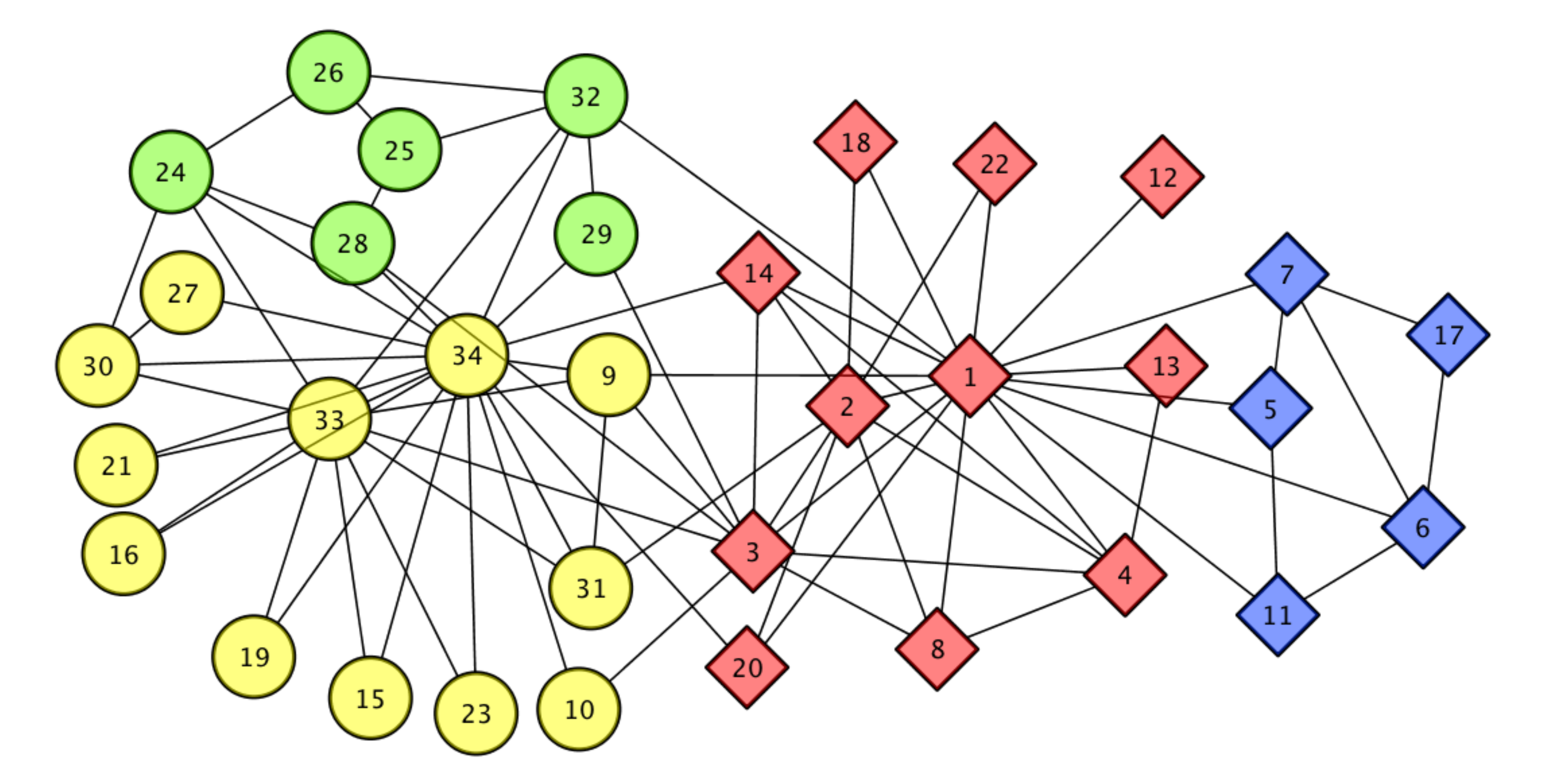}}
\subfloat[]{\includegraphics[width=0.5\textwidth]{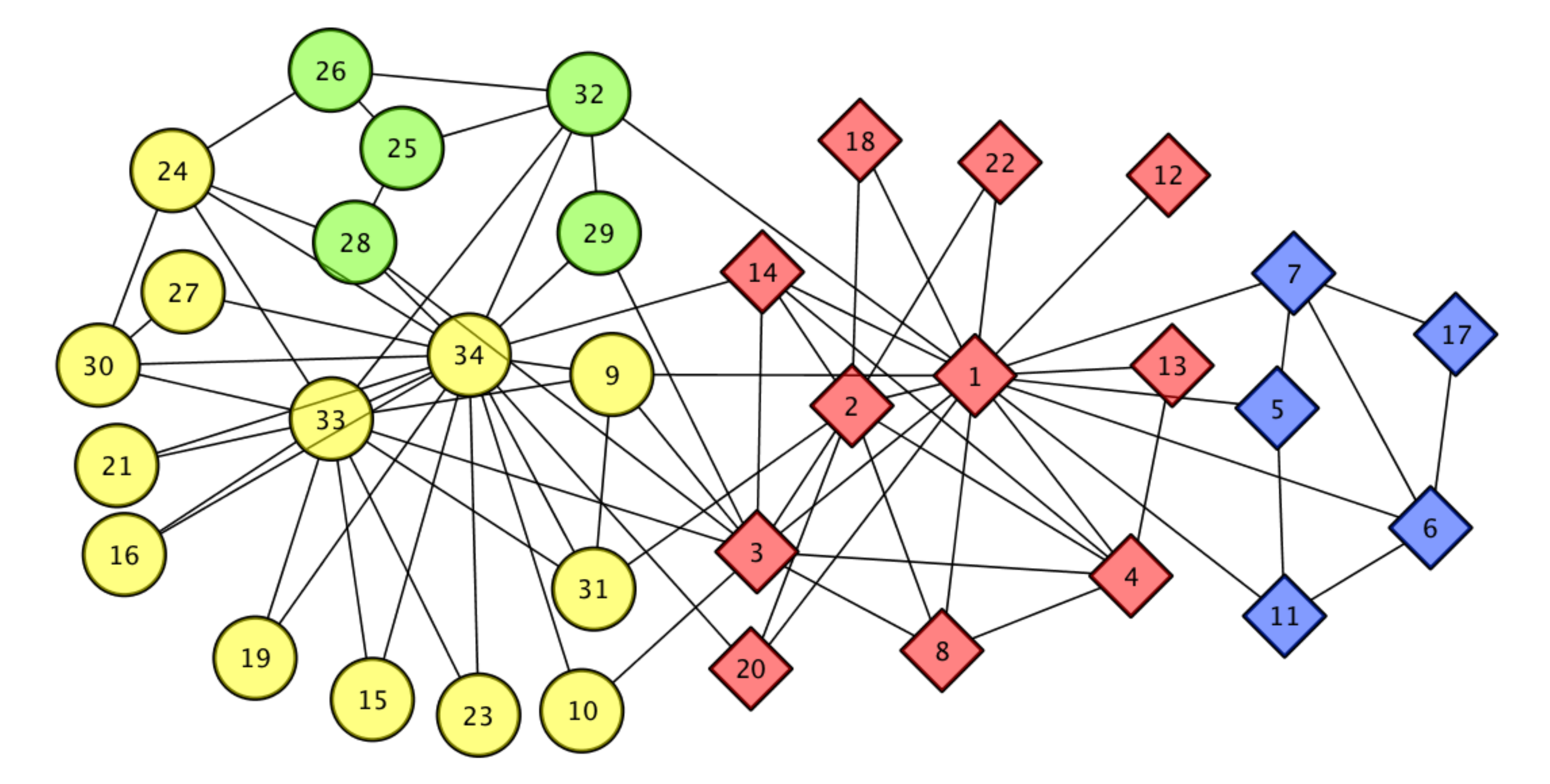}}

\subfloat[]{\includegraphics[width=0.5\textwidth]{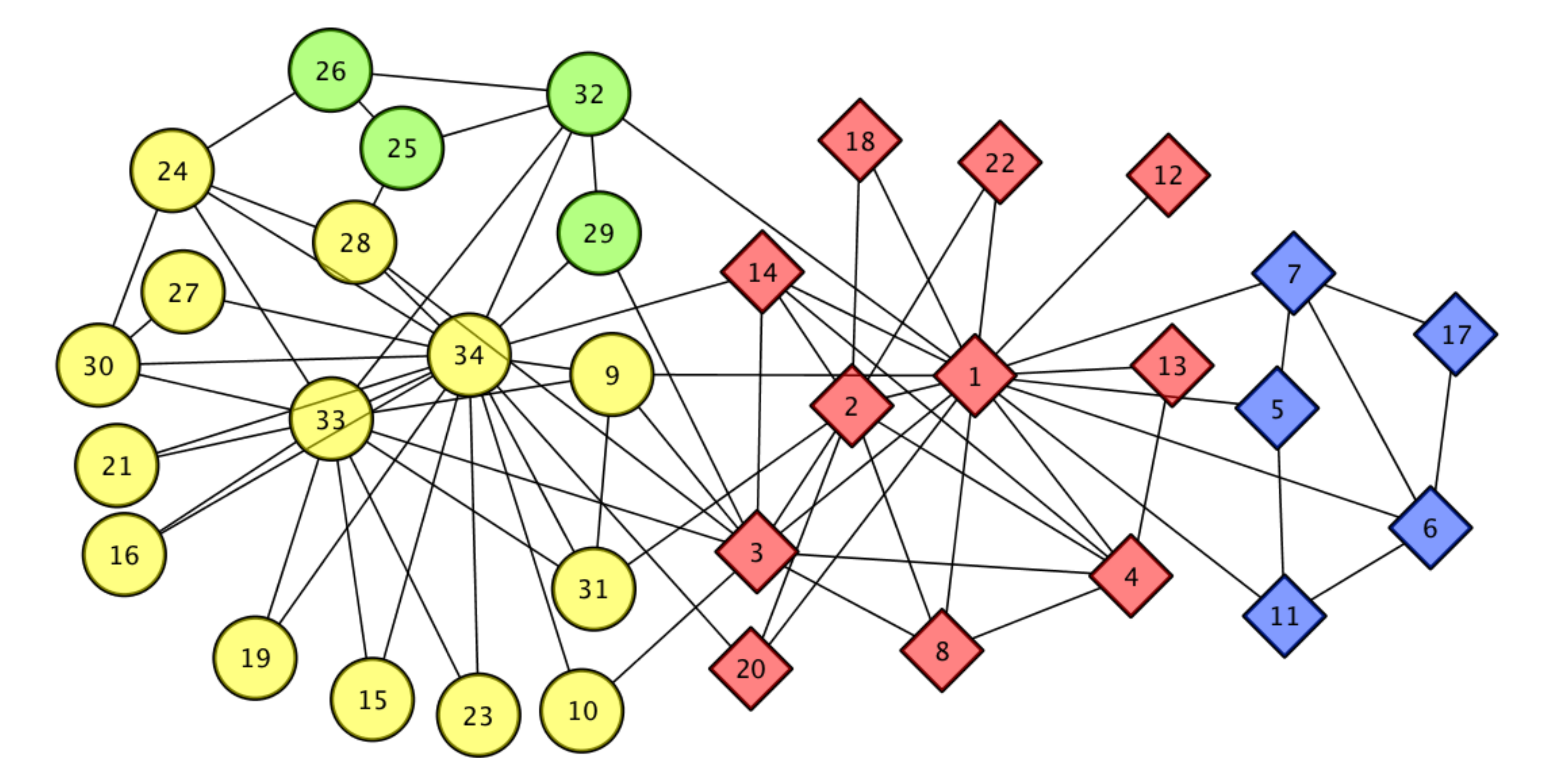}}
\subfloat[]{\includegraphics[width=0.5\textwidth]{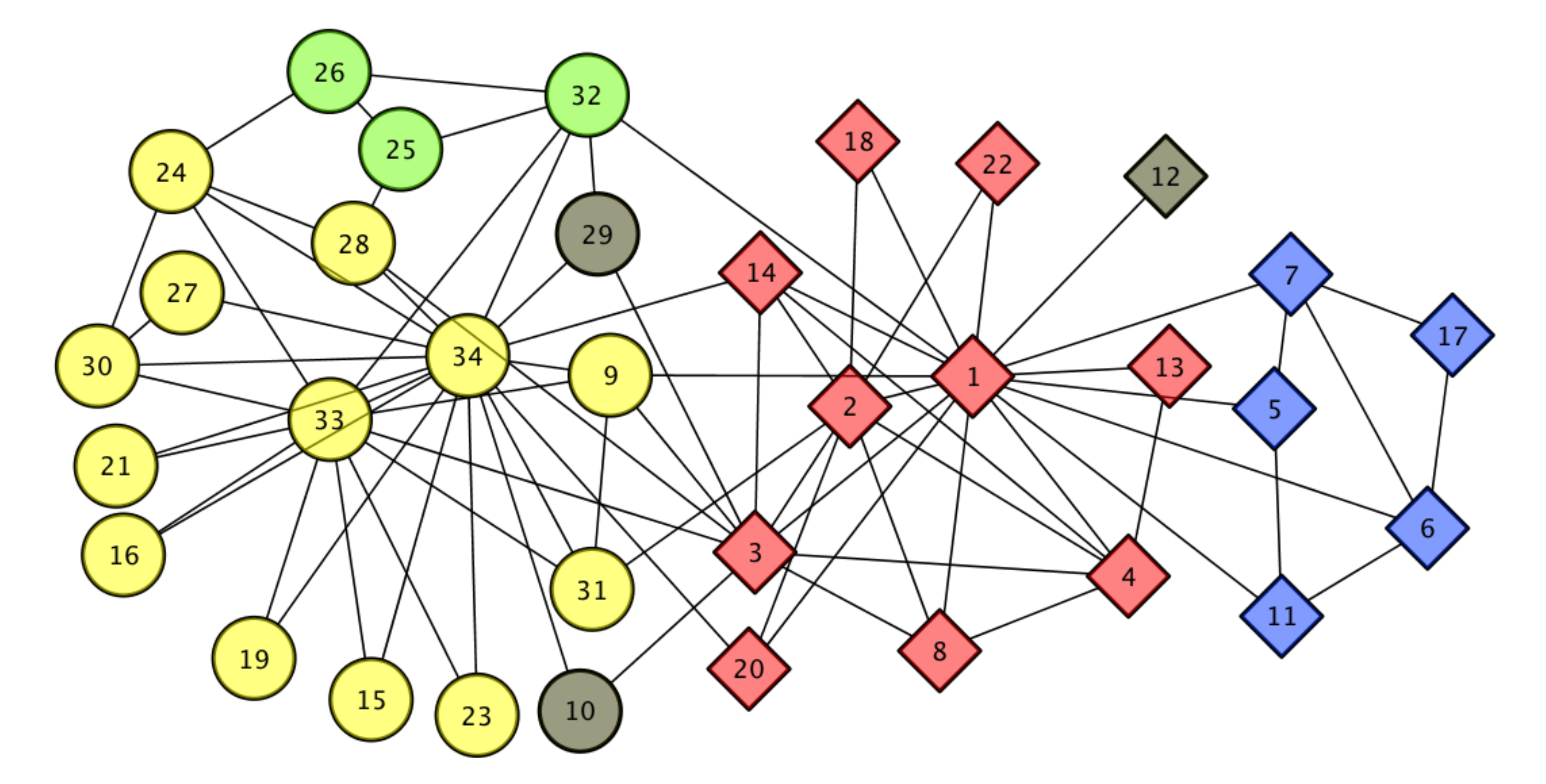}}
\caption{\label{karate} Community detection results of Zachary's Karate Club network. Each color represents one community. The two communities of the ground truth partitioning of the network are represented using circles and diamonds. (a) Partition corresponding to best reported modularity, (b) First reported partition with modularity density value 0.231 \cite{mingming2014}. (c) A recently reported partition with modularity density value 0.235 \cite{charo2016}. (d) Our partition with modularity density value 0.243.}
\end{figure}
Modularity density has been shown to substantially reduce the two problems of modularity mentioned in the introduction while maintaining the general character of communities that modularity finds in practical and synthetic benchmark networks. Using it to analyze the community structure of the well known Zachery's Karate Club~\cite{zachary1977}, a partition with a $Q_{ds}$ value of 0.231 was first reported~\cite{mingming2014}. 
This partition agrees reasonably well with the one thought to maximize $Q$.
These two partitions are shown in Fig.~1(a) and (b). Recently, another partition with a higher $Q_{ds}$ value of 0.235 was reported~\cite{charo2016}. It is shown in Fig.~1(c). 
However, we find a partition with an even higher, potentially the true maximal, $Q_{ds}$ value of 0.243. It is shown in Fig.~1(d).
Finding the network partition that maximizes the metrics $Q$ and, presumably, $Q_{ds}$ is an NP-hard computational problem~\cite{brandes2008}. For the case of $Q$, numerous algorithms have been developed to find good approximate solutions in polynomial time~\cite{newman2004fast,newman2006pnas,Blondel2008,Guimera2004,Medus2005,Duch2005,sun2009}. In this paper, we use variants of an efficient algorithm that was recently introduced in~\cite{trevino2015} to find maximal $Q$ partitions to find partitions that maximize modularity density metrics. This algorithm uses both partitioning and agglomeration, combined with multiple types of Kernigan-Lin-type refinements~\cite{kernigan_lin}, to achieve high-quality partitions.  A similar algorithm was, in principle, used to find the partition in Fig.~1(c) \cite{charo2016}. Our implementation, however, finds the partition with higher $Q_{ds}$ shown in Fig.~1(d). 

Unfortunately, the new partition we find 
reveals an unexpected problem with $Q_{ds}$. Notice that, nodes 10, 12 and 29, which have no direct links between each other, are grouped in the same community. Intuitively, such a partition should not exist. Notice furthermore that, nodes 10, 12 and 29 are somehow special: node 12 is the only node with degree 1 in the network; node 10 has two links which connect to two different communities; nodes 29 has 3 links and each of them connects to a different community. This suggests that letting these three nodes each form a separate single-node community would be an acceptable and better result than putting them together.

To understand the nature of the problem,
consider a partition of a network consisting of $k$ nodes, $a_1, a_2, a_3, \ldots, a_k$, which are isolated from each other, i.e. no links between any pair of these nodes, but may be connected to other nodes in the network, and a set of other nodes that are separated into $m$ communities with sizes $n_1, n_2, n_3, \dots, n_m$. Let the number of links between community $i$ and isolated node $j$ be $l_{ij}$. Then the contributions to value of the SP term, the third term in Eq.~\ref{Qds}, resulting from these links can be calculated. Consider two extreme cases: (1) separating all the $k$ isolated nodes into $k$ communities and (2) merging them into one community. The corresponding contributions to SP term in these two case are, respectively: 
\begin{eqnarray*}
\delta S^{sep} & = & - \sum_{i=1}^{m}\sum_{j=1}^{k}\frac{{l_{ij}}^2}{2mn_i} \\
\delta S^{merge} & = & - \sum_{i=1}^{m}\frac{(\sum_{j=1}^{k}{l_{ij}})^2}{2mkn_i}
\end{eqnarray*}
Since by the RMS-AM inequality~\cite{Gwanyama2004}
\begin{equation*}
\sum_{j=1}^k l_{ij}^2 \geq \frac1k \left( \sum_{j=1}^k l_{ij} \right)^2
\end{equation*}
we have $\delta S^{merge} \geq \delta S^{sep}$.
Thus, the SP term prefers to merge the isolated nodes into one community. 
 
As to the other two terms in $Q_{ds}$, only the communities involving the $n$ isolated nodes can make a difference in their value, since the contribution from the other $m$ communities is the same in both cases. Note that the value of $p_c$ for a community consisting of a single node is not well-defined. Thus, in case (1) when the isolated nodes form separate communities, $Q_{ds}$ is also not well-defined. To fix this problem one can simply define the value of $p_c$ for a single node community, which we refer to as $p_*$. Since it is a density, it is reasonable to expect $p_* \in [0,1]$. Whatever value is chosen for $p_*$, the contribution to the first term in $Q_{ds}$ from communities of the isolated nodes, whether or not they are merged into one community, is always zero because $m_c=0$ for these communities. 
For the second term in $Q_{ds}$, if the nodes are merged, case (2), contributions to its value from
the communities of isolated nodes is 0 because $p_c=0$.
In case (1) though, the contributions to its value depend on the value of $p_*$. If $p_*=0$, then the 
contributions are also 0. However, if $p_*>0$, then the second term would favor case (2), merging the isolated nodes. So, perhaps $p_*$ should be defined to be 0, but even then isolated nodes will tend to be grouped together because of the SP term.
Thus, although the SP term may in some situations solve the problem with modularity of favoring small communities, it also introduces the problem of grouping unlinked nodes into the same community.

\section{Resolution limits of modularity density}

Modularity density, by introducing density coefficients, does solve a well-known RL problem described originally in~\cite{fortunato2007}. As shown in Fig. 2, modularity fails to resolve pairs of cliques, i.e. fully connected sets of nodes, in certain configurations. In two cases shown, cliques are only connected by a single link and, thus, can be expected to form independent communities. However, if modularity density is used instead, then the cliques are resolved in these two cases. Thus, using modularity density does significantly address this RL problem. In these two examples though, the cliques in each pair both have the same size. If instead they have unequal sizes, then the results are more complicated. 

\begin{figure}
\centering
\includegraphics[width=1.0\textwidth]{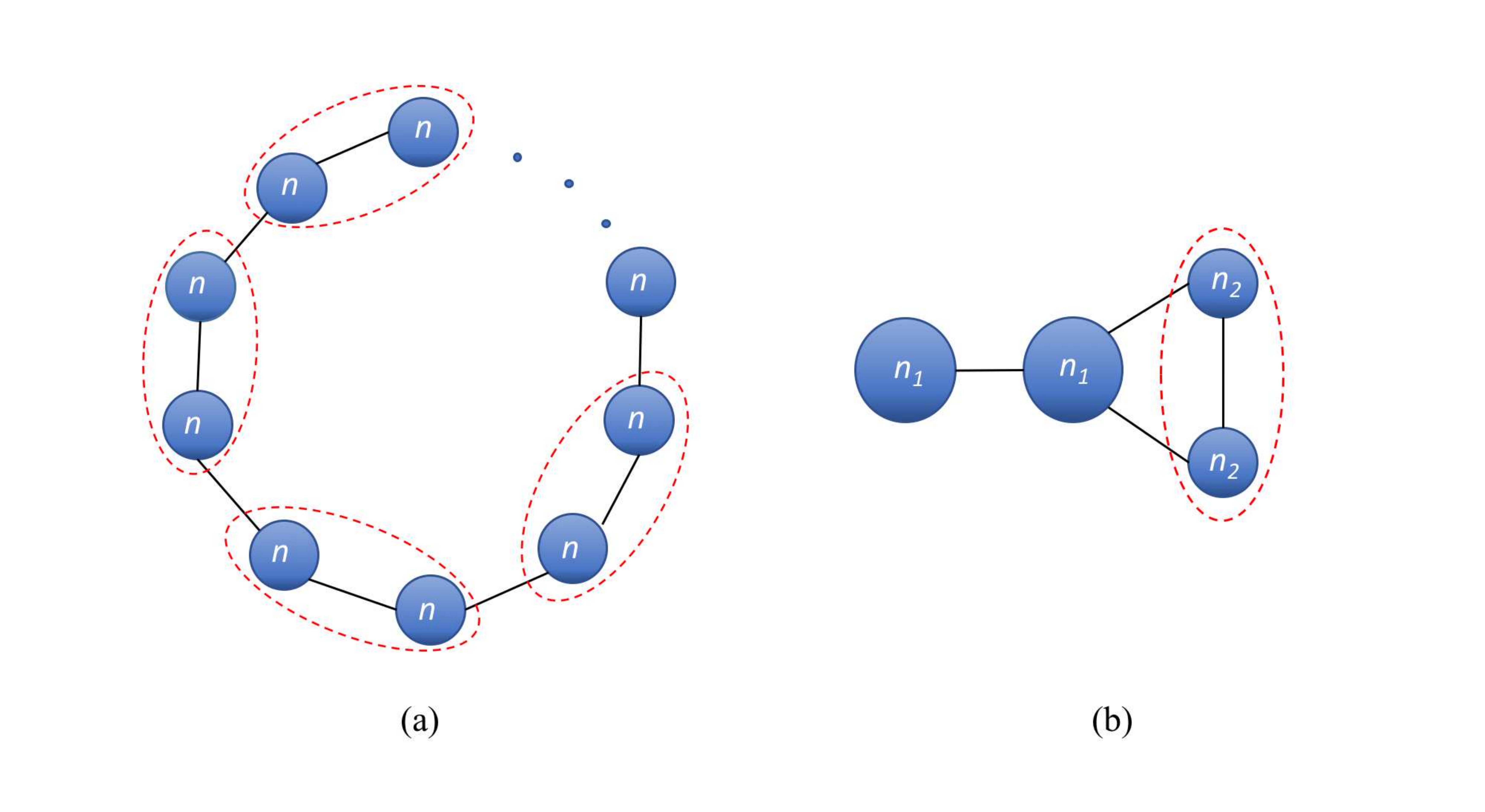}
\caption{\label{resolution limit} Resolution limit examples of modularity. 
(a) A ring of cliques, each of size $n$, with each clique connected by a single link. (b) Two pairs of cliques of sizes $n_1$ and $n_2$, connected by single links as shown. When the number of cliques in (a) is large and when $n_1/n_2$ is large in (b), modularity fails to resolve the pairs of cliques circled by red ellipses. Modularity density, however, resolves these clique pairs. }
\end{figure}

\begin{figure}
\centering
\subfloat[]{\includegraphics[width=0.38\textwidth]{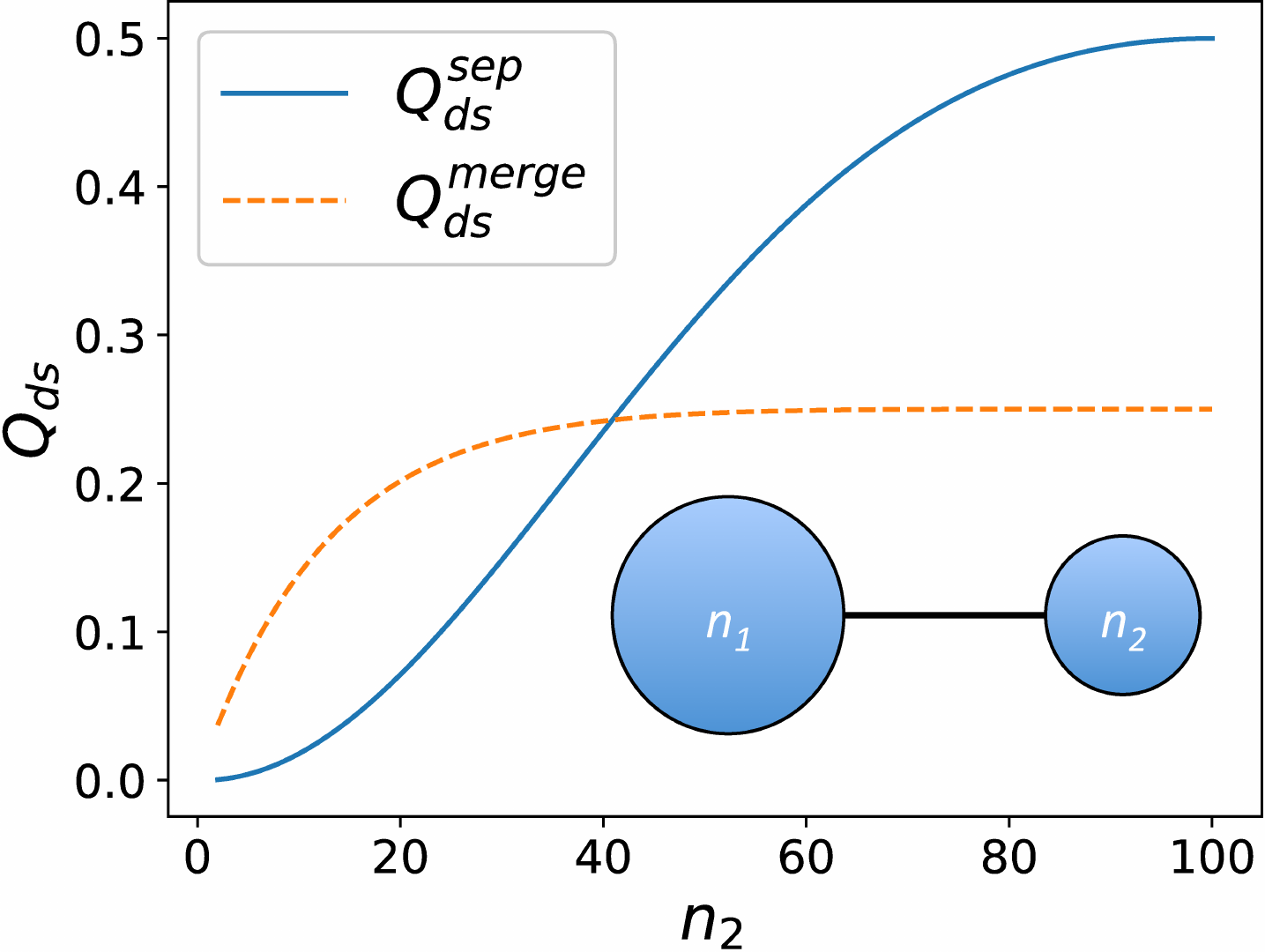}}
\subfloat[] {\includegraphics[width=0.301\textwidth]{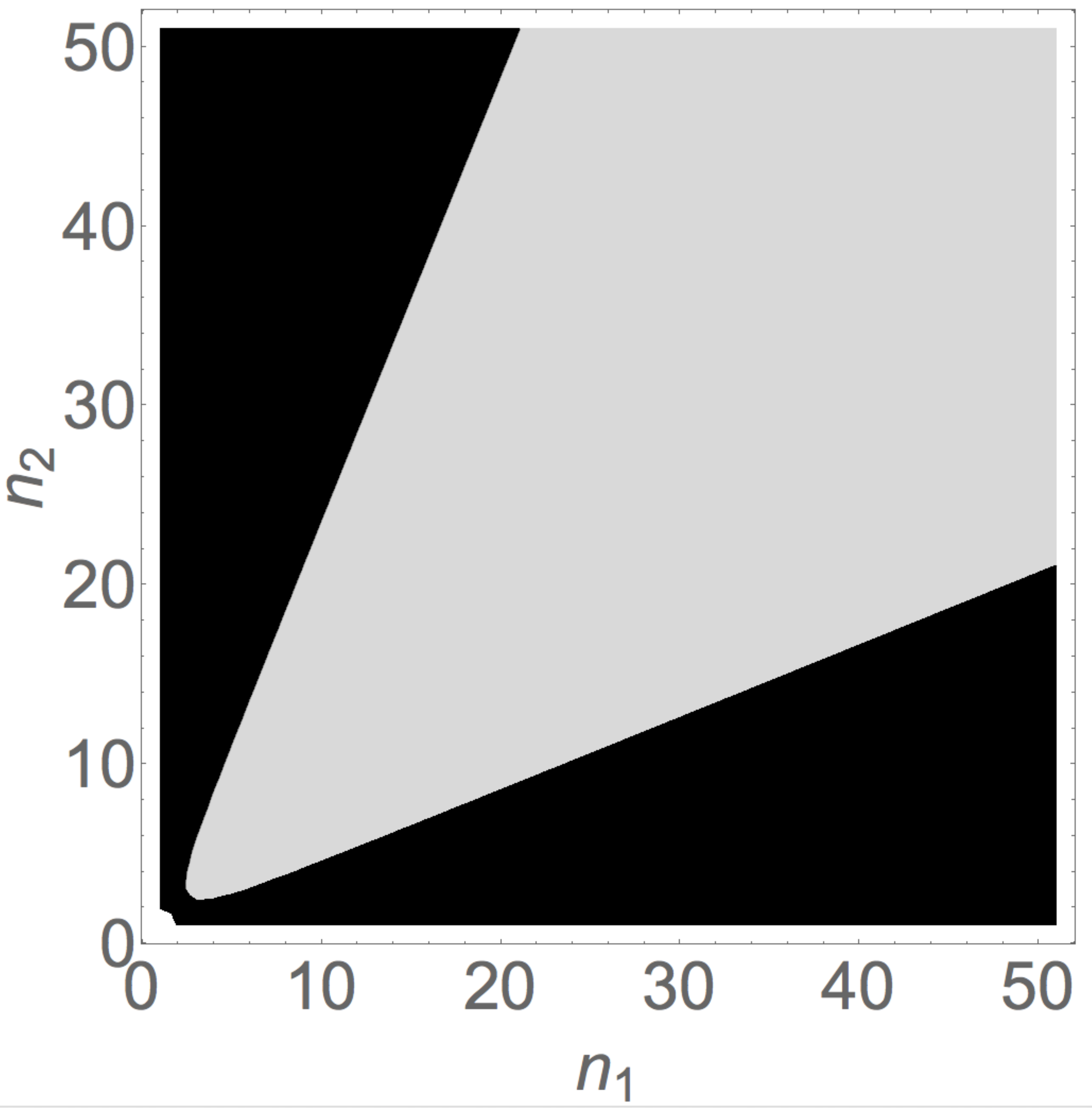}}
\subfloat[]{\includegraphics[width=0.31\textwidth]{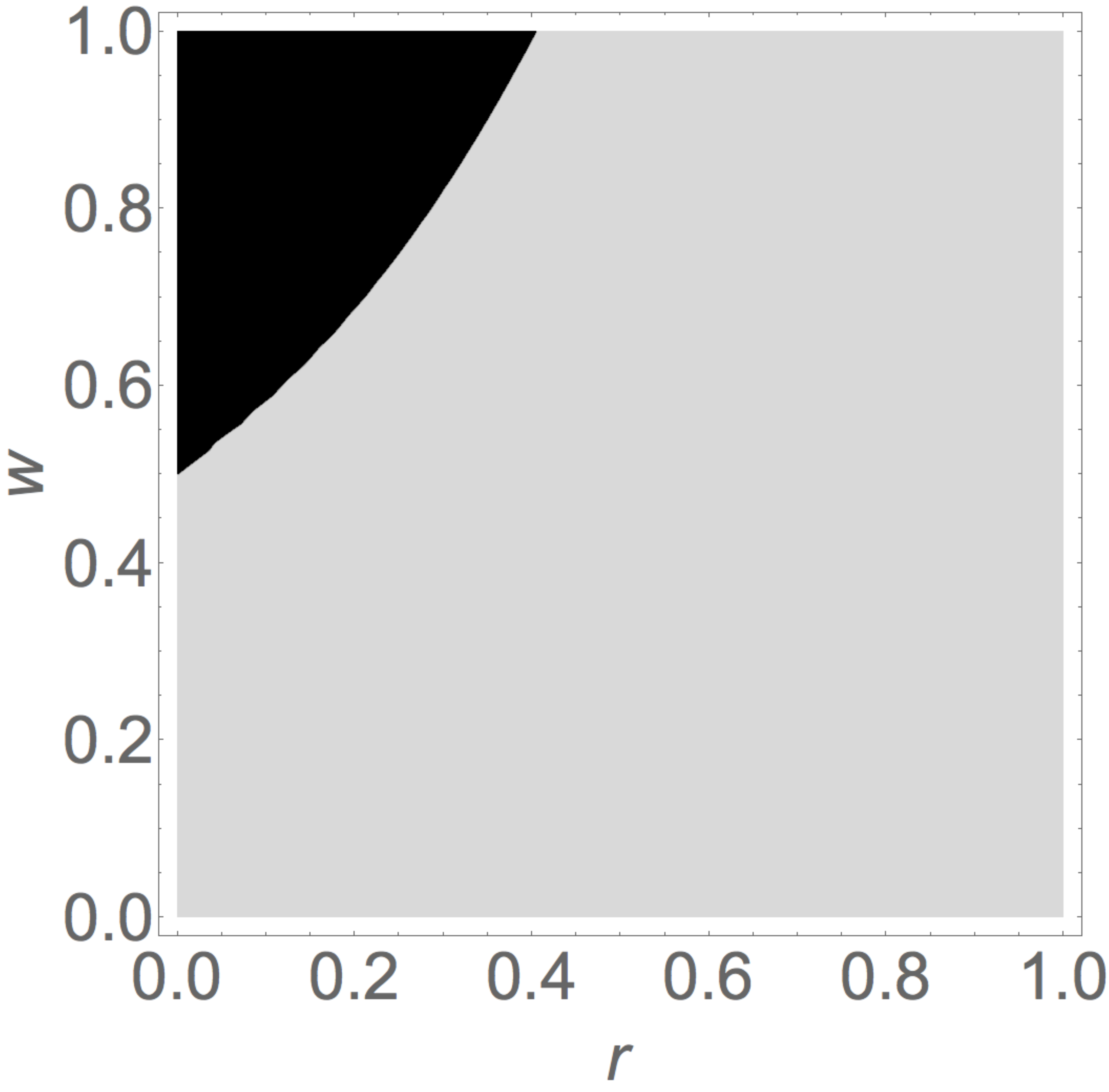}}
\caption{\label{twoclique} Resolution limit examples of modularity density. (a) Modularity density of a network consisting of two cliques of sizes $n_1$ and $n_2$ connected by single link (as shown in the inset) when then cliques are merged $Q_{ds}^{merge}$ and separated $Q_{ds}^{sep}$. Here $n_1 = 100$ (fixed) and $n_2$ is varied. 
(b) Ability of modularity density to resolve the cliques in a network consisting of just two cliques with sizes $n_1$ and $n_2$ connected by a single link as a function of clique sizes.
The cliques are resolved only in the light gray region; they are unresolved in the black region.
(c) Ability of modularity density to resolve two cliques with sizes $n_1$ and $n_2$ within a larger network as a function of the relative size of the cliques $r$ and the fraction of total links in the network contained within the two cliques $w$.
Again, the cliques are resolved only in the light gray region; they are unresolved in the black region.
}
\end{figure}

Consider the particularly simple example of two cliques with sizes $n_1$ and $n_2$, with just one link connecting them, as shown in the inset of Fig.~\ref{twoclique}(a), 
and define the relative size of the cliques as $r=n_2/n_1$. 
Let $Q_{ds}^{sep}$ and and $Q_{ds}^{merge}$ denote the modularity density when the two cliques are assigned to two separate communities and when the two cliques are merged into a single community, respectively. 
Figure~\ref{twoclique}(a) shows results for $n_1=100$ and varying $n_2$ that reveal how the relative sizes of the two cliques affect the ability of $Q_{ds}$ to resolve them.
When $n_2 \lessapprox 40$, i.e. $r \lessapprox 0.4$, $Q_{ds}^{merge} > Q_{ds}^{sep}$ and modularity density does not resolve the two cliques.
Figure~\ref{twoclique}(b), similarly, shows the values of $n_1$ and $n_2$ for which modularity density does not resolve the cliques when they are connected by a single link.

The critical value of $r$, below which $Q_{ds}$ fails to resolve a pair of cliques, can also be estimated assuming that the cliques are large and the number of links connecting them is small.  In this case,  $m_c = n_c(n_c - 1)/2 \approx n_c^2/2$ and $2m_c + e_c \approx 2m_c$ for each clique. Furthermore, $p_{cc'} \ll 1$ and the SP term can be ignored. Then, after simplification we get
\begin{eqnarray*}
\Delta Q_{ds} & \equiv & Q_{ds}^{merge} - Q_{ds}^{sep} \approx 2r(1+r^2)/(1+r)^4-2r^2/(1+r^2)^2
\end{eqnarray*}
The equation $\Delta Q_{ds} = 0$ has two real roots that indicate that modularity density fails to resolve the cliques ($\Delta Q_{ds} > 0$) if $r \lessapprox 0.405$ or $r \gtrapprox 2.470$. This result is independent of system size (for large $n_1$ and $n_2$) and  agrees well with the results shown in Figs.~\ref{twoclique}(a) and (b).

More generally, if the two cliques are part of an unspecified larger network, then the ability of $Q_{ds}$ to resolve them is a function of $r$ and of the fraction of the network's links that are contained within the two cliques $w = (m_1+m_2)/m$ can also be evaluated. Here $m_1$ and $m_2$ are the number of links contained within the cliques. In this case, the black region of Fig.~\ref{twoclique}(c) indicates where $\Delta Q_{ds} >0$ and the cliques are not resolved.
Hence, modularity density fails to resolve two cliques when the sizes of cliques are not balanced, specifically when the small clique is smaller than about 0.4 of the large clique size, and the links contained within the two cliques account for more than about half of the total number of links in the network. 
Considering again the two cases shown in Fig.~\ref{resolution limit}, Fig.~\ref{twoclique}(c) indicates that modularity density is able to resolve the cliques in each pair because 
they have equal sizes $r = 1$, but that it potentially would not if $r \lessapprox 0.4$.
Cliques are, of course, an extreme form of dense community, and we have discussed only the case when networks consist only of cliques connected by single links, but our conclusions about the RL of $Q_{ds}$ remain approximately true if the the communities are dense, but not cliques, and if they are connected by a sufficiently small number of links or no links at all.

Thus, although modularity density does significantly mitigate the resolution limit problem of modularity in many situations, it does not solve the problem completely. In the situations discussed above, modularity density still suffers from a resolution problem. 
Nevertheless, modularity density is still a good metric if the network is within its domain of applicability, such as the light gray regions in Figs.~\ref{twoclique}(b) and (c). 

\section{An improved metric: Excess modularity density}
\begin{figure}
\centering
\includegraphics[width=0.6\textwidth]{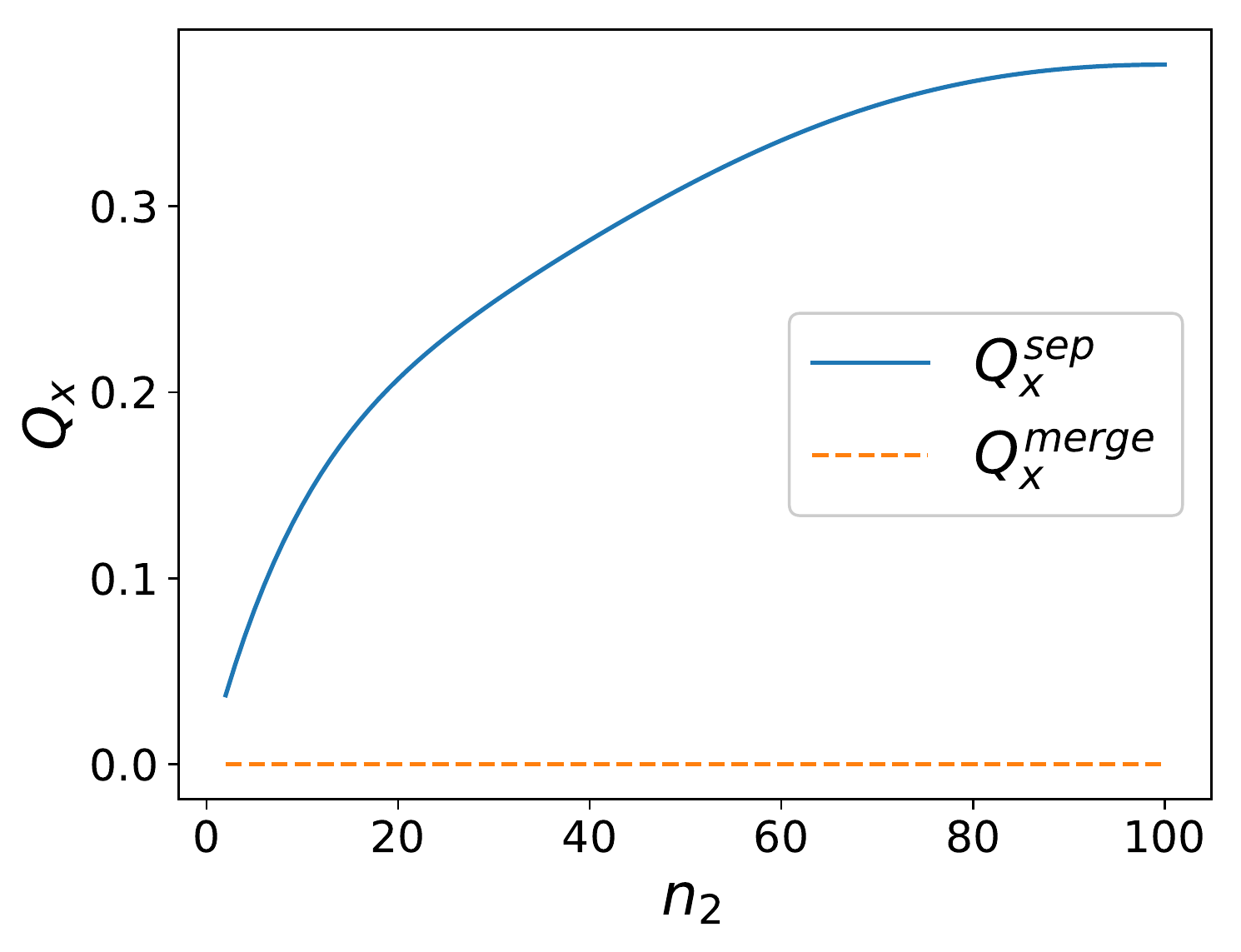}
\caption{\label{qx_twoclique} Absence of a resolution limit for excess modularity density for a two-clique network. The first clique has $n_1 = 100$ nodes, the size of the second clique $n_2$ is varied, and one link connects them. $Q_{x}$ is always larger when the cliques are separated rather then merged.}
\end{figure}

\begin{figure}
\centering
\includegraphics[width=0.9\textwidth]{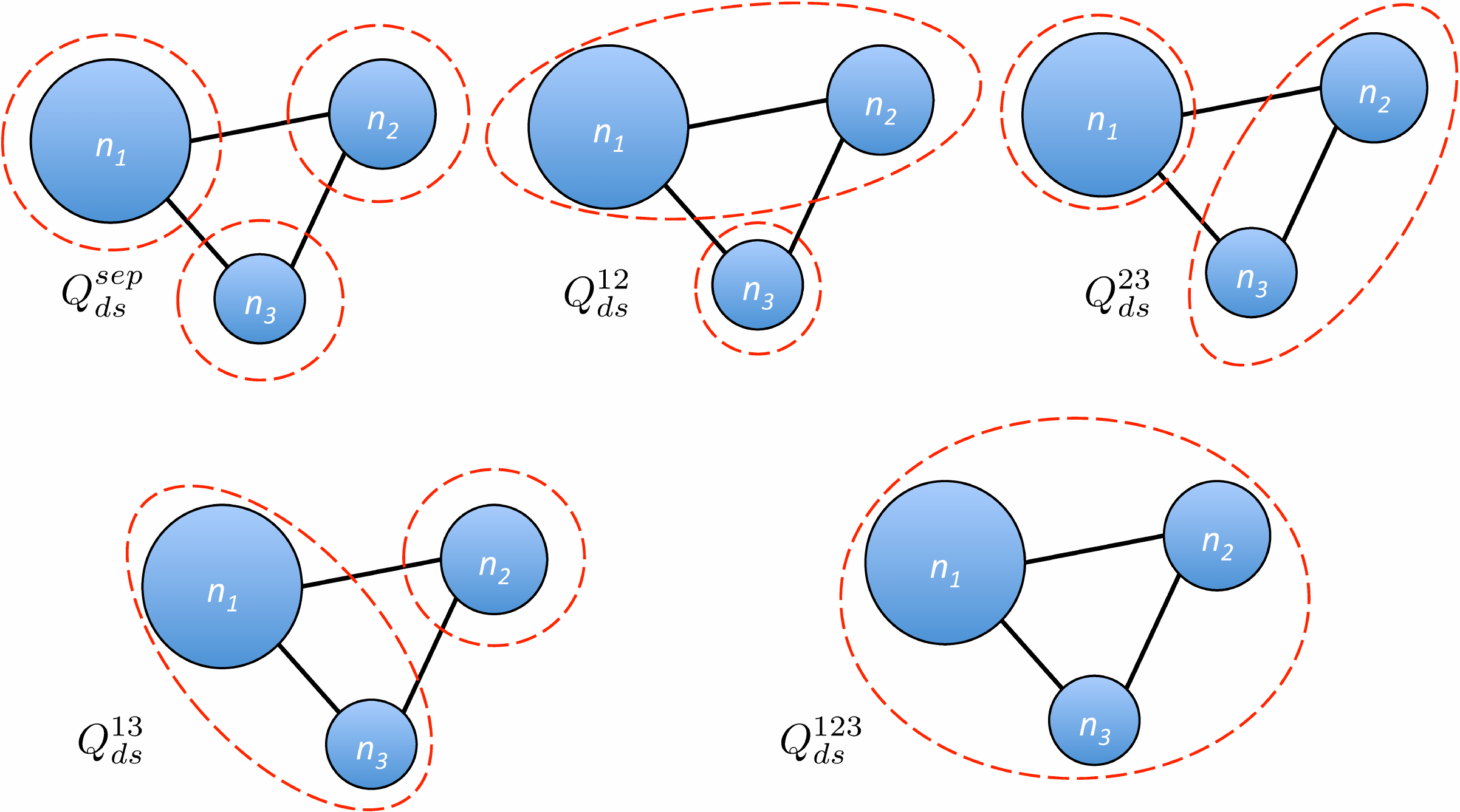}
\caption{\label{threeclique} Partitions of a three-clique network. All possible partitions of the three-clique network without splitting a clique are shown. In each case cliques enclosed by the dashed lines are grouped together.}
\end{figure}

Because of the inclusion of the link density $p_c$ in the definition of modularity density, Eq.~\ref{Qds}, when all nodes are grouped together in one community, except in trivial cases, the second term is smaller than the first. 
Because of this, even in the absence of any modular structure in the network, $Q_{ds}$ will be positive. 
Whereas, the value of modularity $Q$ for this case is zero. 
In order to fix this issue, we propose using a modified density metric for the quality of community structure. Our metric replaces the community link density $p_c$ in the definition of $Q_{ds}$ by a rescaled link density 
\begin{equation}
p_{c}^\prime = p_c - \frac{2m}{N(N-1)}
\end{equation} 
where $m$ and $N$ are the total number of links and nodes of the network, respectively.
$p_c^{\prime}$ measures the excess link density inside a community by subtracting the global link density from $p_c$.
This is intuitively attractive because $p_c^{\prime}$ measures the link density in a community above and beyond the global average. It also eliminates the problem of measuring a positive non-zero modularity density even in the absence of any modular structures.
We also exclude the SP term to avoid the problems caused by it that were discussed in Sec.~\ref{SP}.  We denote this new metric by $Q_{x}$ and refer to it as {\it excess modularity density}
\begin{eqnarray}
Q_{x} = \sum_{c \in C}\left[\frac{m_c}{m} p_{c}^\prime -\left(\frac{2 m_c + e_c}{2m} p_{c}^\prime \right)^2 \right]
\end{eqnarray}
The partition that maximizes $Q_{x}$ corresponds to the community structure. An added advantage of excluding the SP term is that it makes finding the maximal partition easier computationally.

\begin{figure}
\centering
\subfloat[]{\includegraphics[width=.31\textwidth]{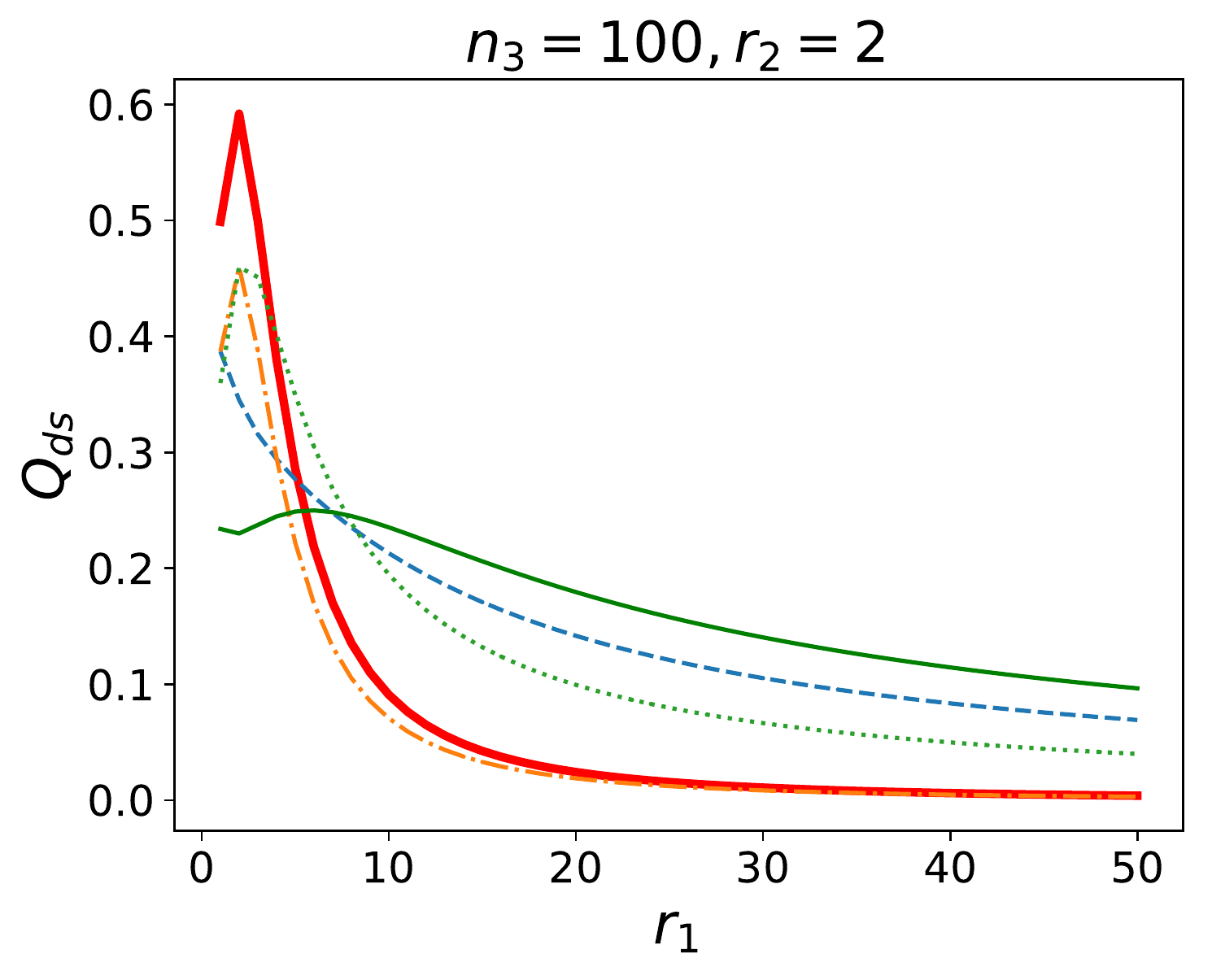}}
\subfloat[]{\includegraphics[width=.31\textwidth]{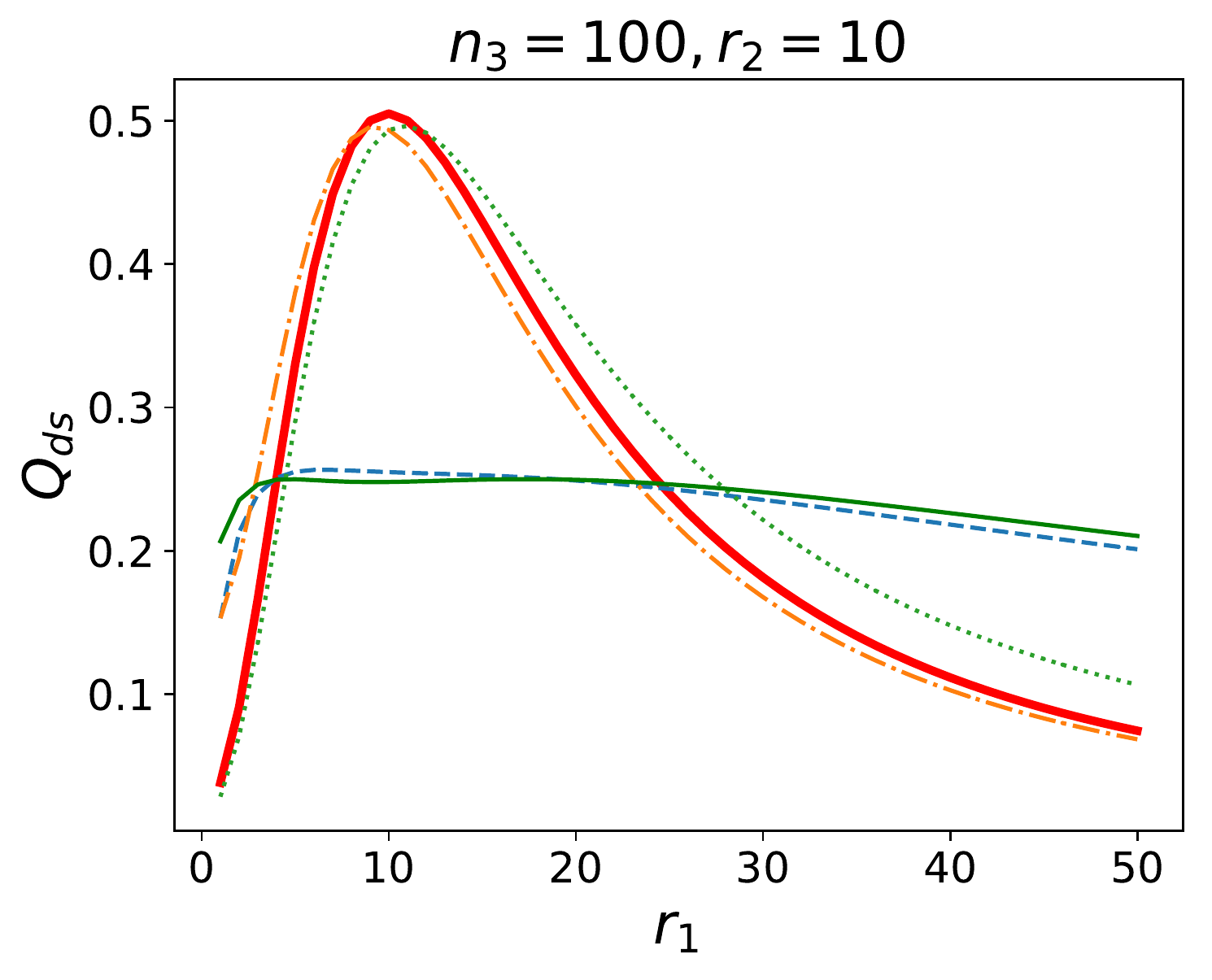}}
\subfloat[]{\includegraphics[width=.38\textwidth]{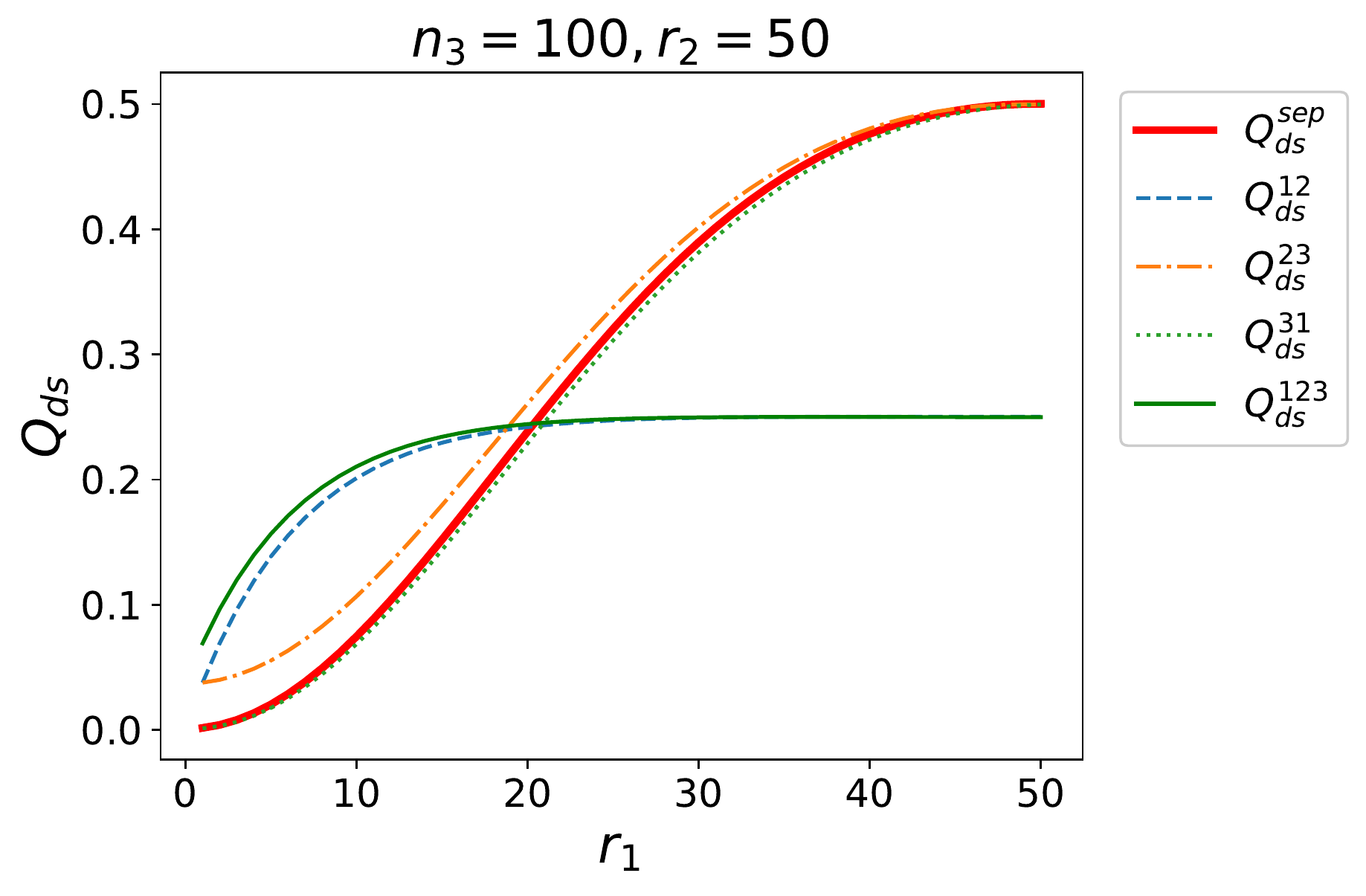}}\\
\subfloat[]{\includegraphics[width=.31\textwidth]{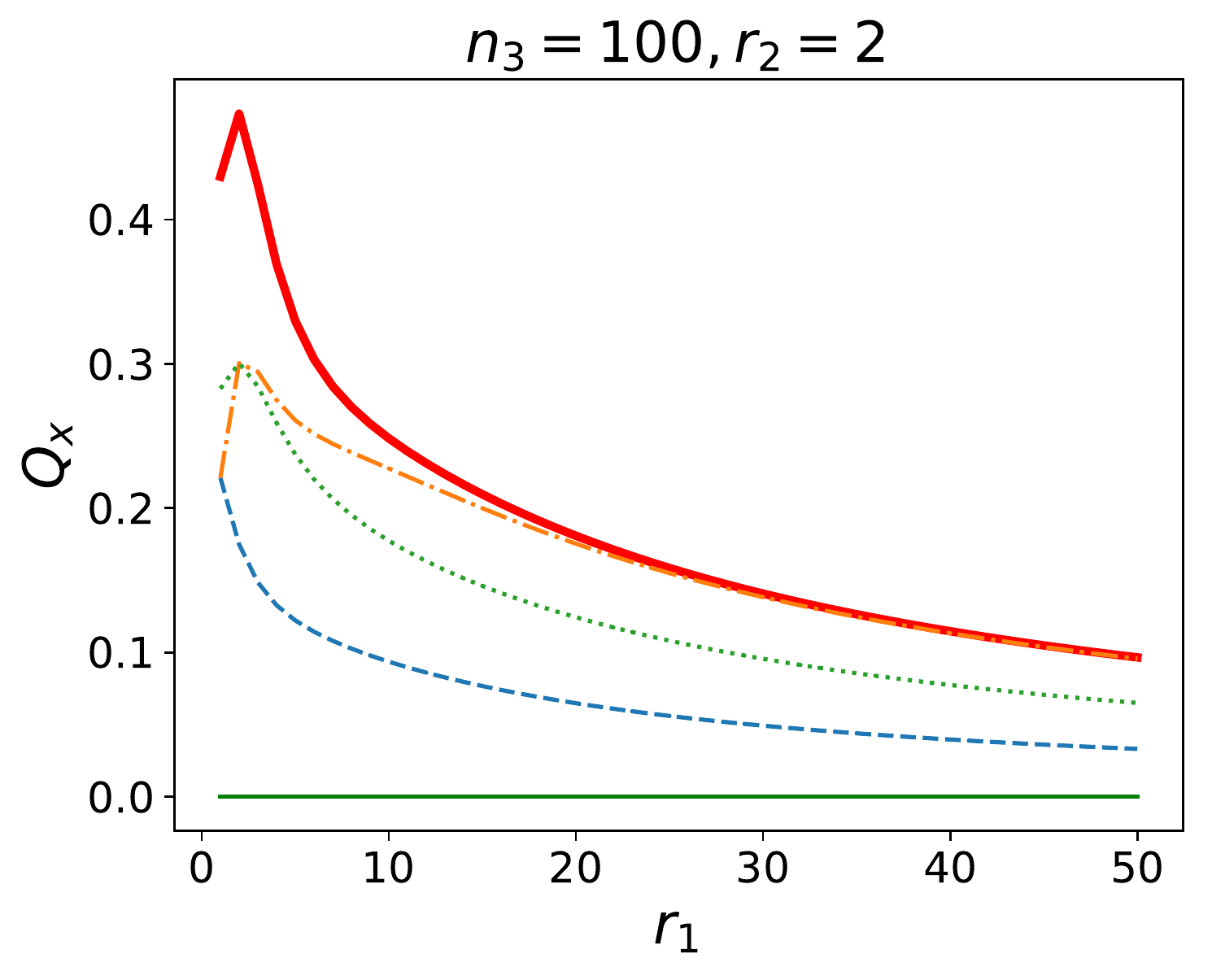}}
\subfloat[]{\includegraphics[width=.31\textwidth]{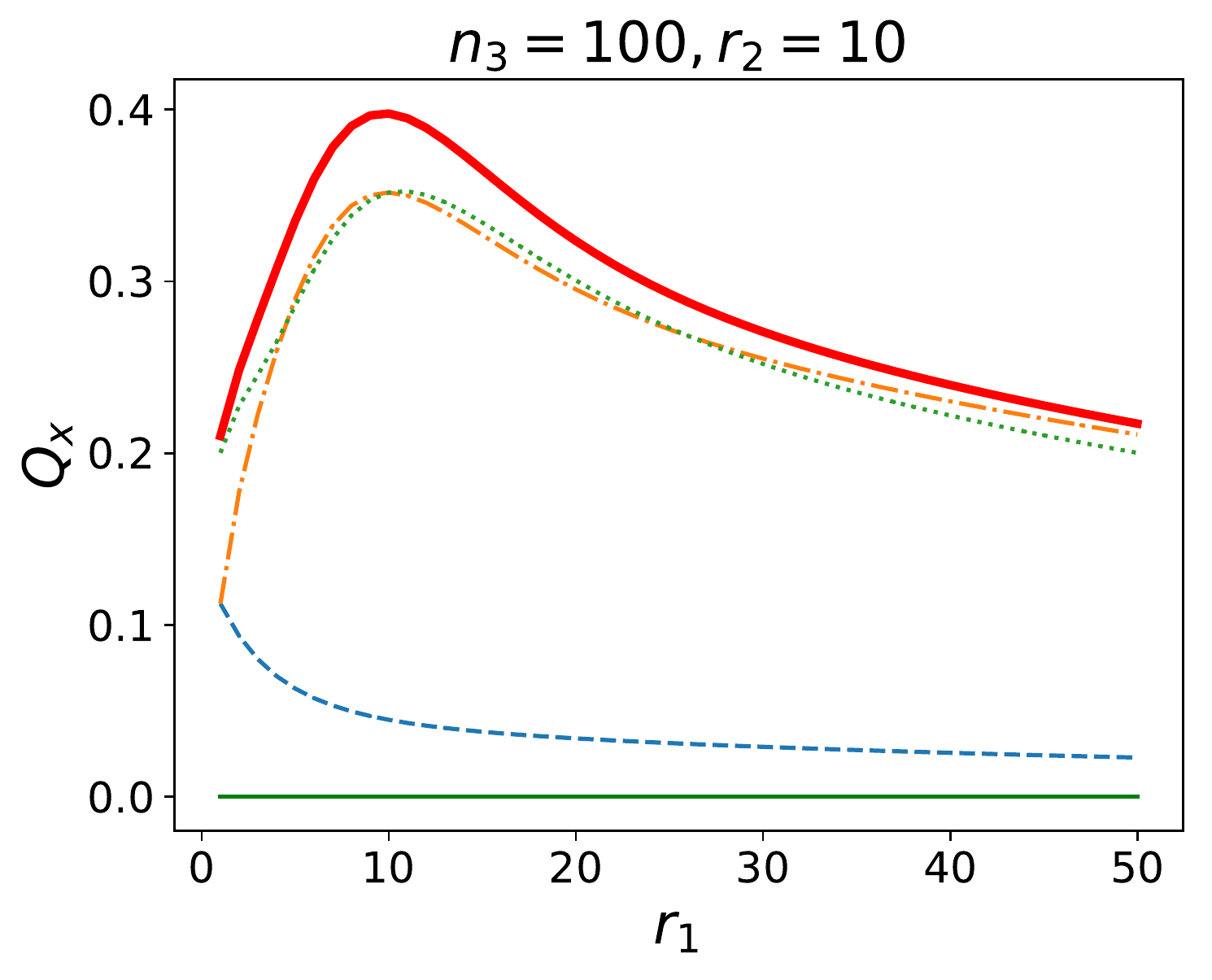}}
\subfloat[]{\includegraphics[width=.38\textwidth]{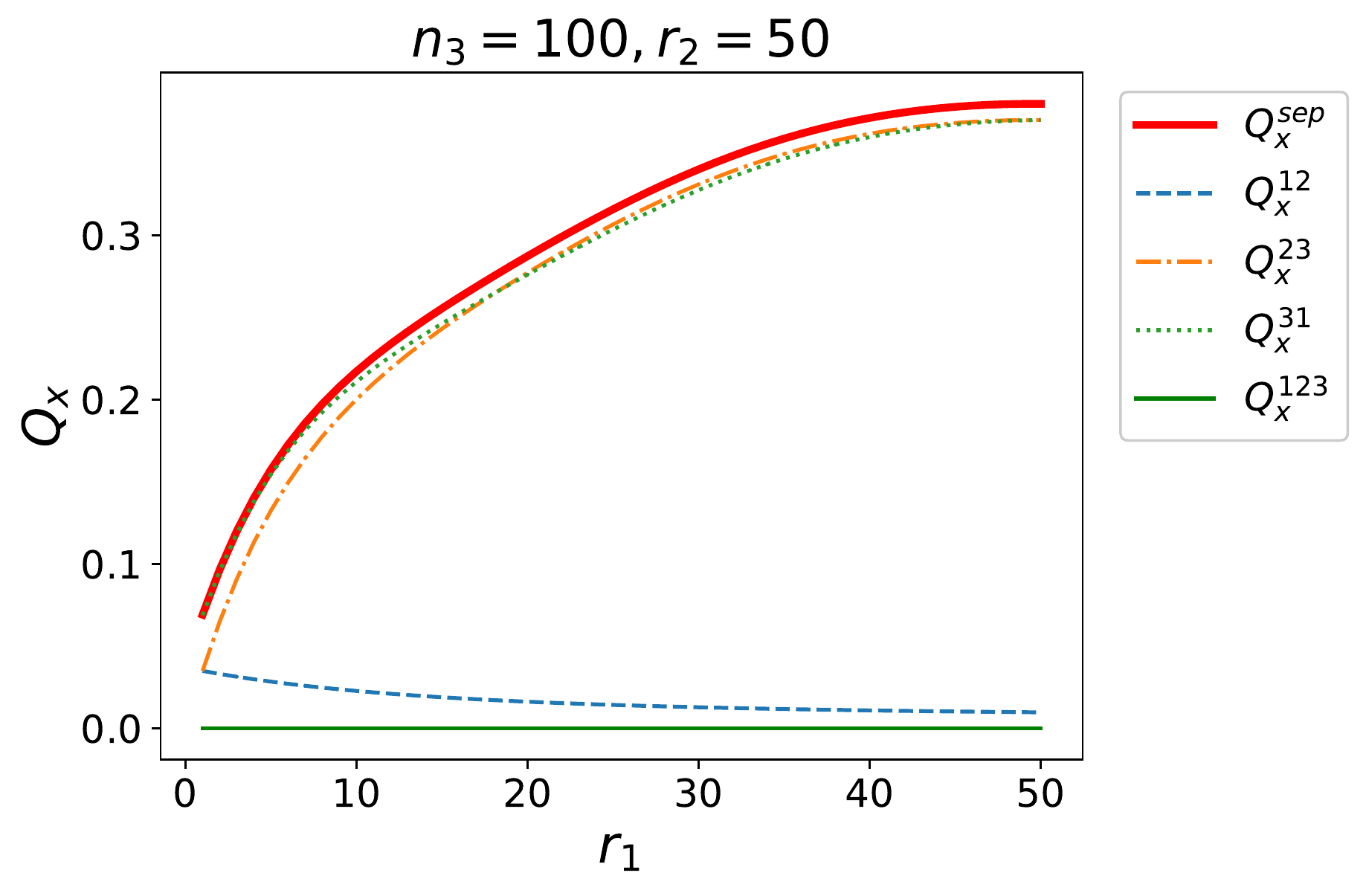}}
\caption{\label{qds_qx} Comparison of modularity density measures for partitioning a network of three cliques of different sizes. (a)-(c) Modularity density $Q_{ds}$ for different partitions and different relative sizes of the three cliques given by $r_1$ and $r_2$. (d)-(f) Excess modularity density $Q_{x}$ for the same partitions and same clique sizes. In all cases, $n_3=100$ and each pair of cliques is connected by a single link.}
\end{figure}

To fully define $Q_x$, a value of the link density for single node communities $p_*$ must also be defined. Given the considerations of Sec.~2, an appropriate value of $p_*$ should not cause disconnected nodes to be grouped together. Consider a set of $n$ isolated nodes that have no common links, but may be connected to the rest of the network. The values of $Q_x$ when the the isolated nodes are merged $Q_x^{merge}$ and when each node is assigned a separate community $Q_x^{sep}$, assuming the same community structure in the rest of the network in both cases, can be written as
\begin{eqnarray*}
Q_x^{merge} & = & -\left[\frac{\sum_{i=1}^{n} d_i}{m} \left(0 - p\right)\right]^2 + Q_x^{rest} = - \frac{p^2}{4m^2} \left(\sum_{i=1}^{n} d_i\right)^2 + Q_x^{rest},\\
Q_x^{sep} & = & \sum_{i=1}^{n} \left\{- \left[\frac{d_i}{2m} \left(p_* - p\right)\right]^2\right\} + Q_x^{rest} = - \frac{(p_* - p)^2}{4m^2} \sum_{i=1}^{n} d_i^2 + Q_x^{rest}.
\end{eqnarray*}
Here $d_i$ is the degree of isolated node $i$, $p = 2m/[N(N-1)]$ is the density of links in the total network, and the $Q_x^{rest}$ is the contribution to $Q_x$ from the remaining portion of the network. Since $\left(\sum_{i=1}^{n} d_i\right)^2 \geq \sum_{i=1}^{n} d_i^2$, if $p_* = 0$ then $Q_x^{sep} \geq Q_x^{merge}$, which is not necessarily true for any other choice of $p_*$. Hence, we define $p_* = 0$.

To demonstrate the efficacy of using $Q_x$, consider again the problem of two cliques of different sizes that $Q_{ds}$ fails to resolve if the sizes are too different, Figure~\ref{twoclique}(a). 
Figure~\ref{qx_twoclique} shows the analogous results using $Q_x$. As $Q_x^{sep}>Q_x^{merge}$ for all values of $n_2$, there is no resolution limit problem using $Q_x$ in this case.
In fact, there is no resolution limit problem using $Q_x$ in any range of the more general two-clique problem shown in Figs.~\ref{twoclique}(b) and (c), even if the sizes of the cliques are extremely different.

\begin{figure}
\centering
\subfloat[]{\includegraphics[width=0.9\textwidth]{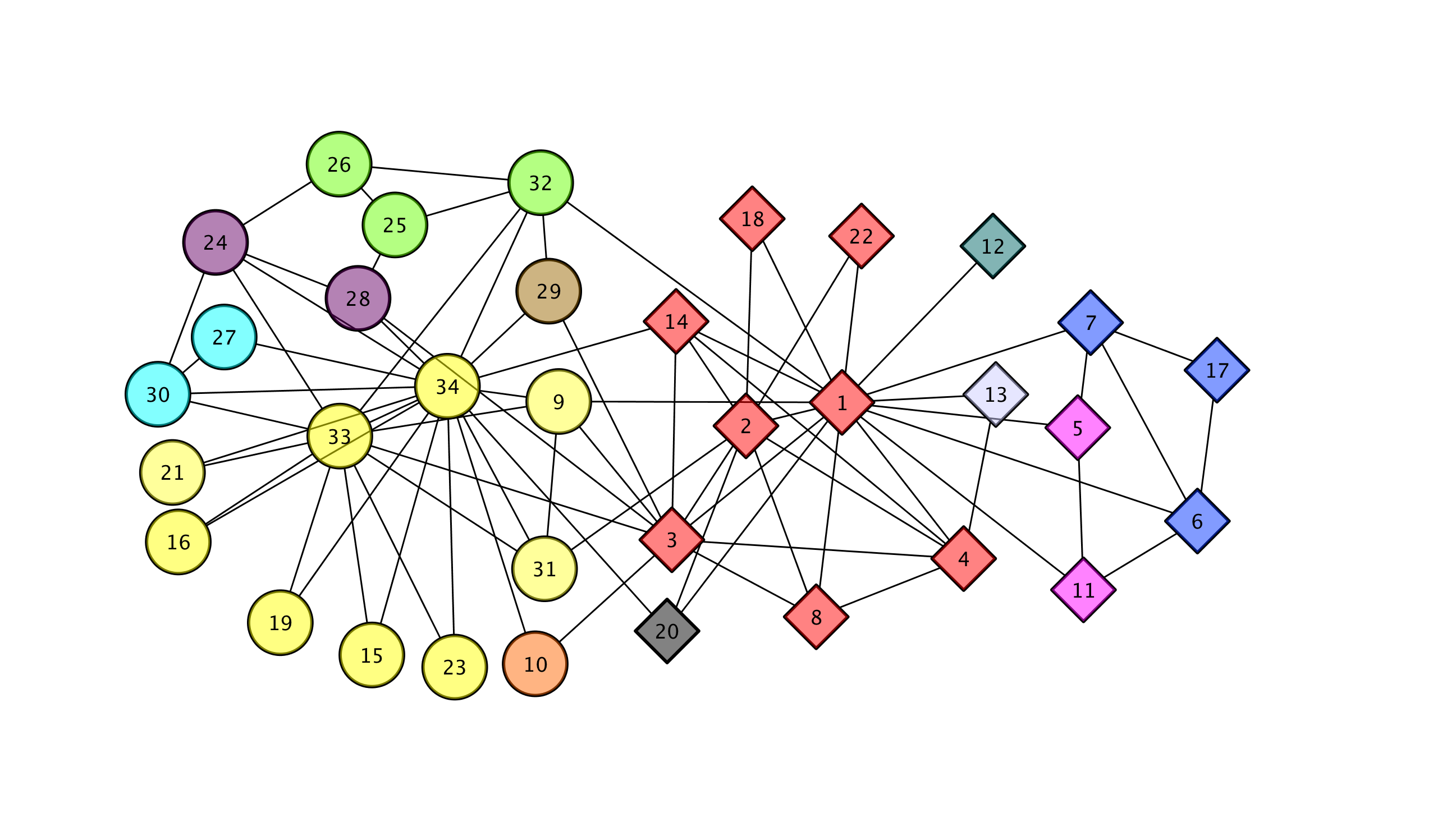}}\\
\subfloat[]{\includegraphics[width=1.0\textwidth]{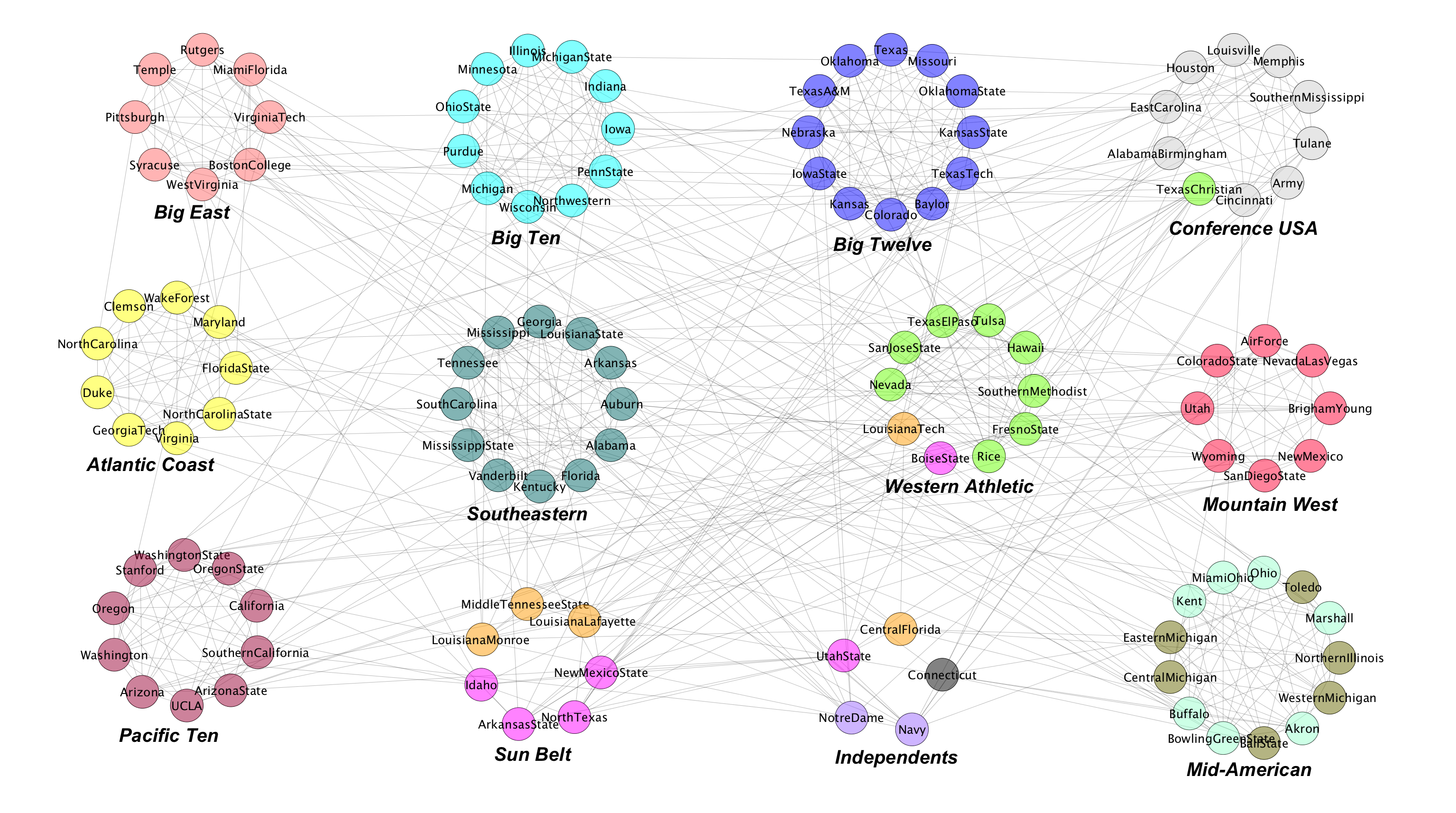}}
\caption{\label{empirical} Community detection in empirical networks using $Q_x$. Nodes belonging to the same community are indicated by the same color. (a) The partitioning result of Zachary's Karate Club network with $Q_x$=0.227. The two communities of the ground truth partitioning of the network are indicated by circles and diamonds. (b) The partitioning result of the American college football network with $Q_x$=0.467. The 12 communities of the ground truth partitioning of the network are the conferences indicated by the layout of the 12 circles of nodes.}
\end{figure}

Next, we consider a more general case of three cliques with sizes $n_1$,$n_2$, and $n_3$, each pair of which is connected by a single link. Let $r_1 = n_1/n_3$, $r_2 = n_2/n_3$. Figure ~\ref{threeclique} shows the five possible ways to partition the network without dividing the cliques. 
Figure~\ref{qds_qx} shows the value of modularity density and excess modularity density for an example set of relative clique sizes for each of these five possible partitions.
Figures~\ref{qds_qx}(a)-(c) show the value of modularity density,
while Figs.~\ref{qds_qx}(a)-(c) show the value of excess modularity density.
Note that $Q_{ds}^{sep}$ is not always the highest among the various partitions, but that $Q_{x}^{sep}$ is always the highest of any partition. 
Thus, modularity density can fail to resolve the three cliques, but excess modularity density is always able to do so in the cases considered.

Although we have only shown the limitation of $Q_{ds}$ in two-clique and three-clique examples here, we believe that similar issues would be encountered when analyzing more complex network structures. $Q_{x}$ can help eliminate these problems to a great extent but it may also have limitations and we have not shown that it is guaranteed to work in the most general case. In fact in some extreme cases, we expect $Q_x$ will fail to resolve communities. Consider, for example, the two-clique network of Figure~\ref{twoclique} but embedded in a larger network. By adding more communities to the network that are loosely connected to the two cliques the global link density $p$ can be made to systematically approach zero. In this limit, $p_{c}^\prime \rightarrow p_{c}$ and consequently $Q_{x} \rightarrow Q_{ds}$ without the {\it SP} term. But even in this extreme example, $Q_x$ is at least as good a metric as $Q_{ds}$. Many networks of interest as not as extreme and, so, $Q_{x}$ can be used as an improved metric that reduces the resolution limit problems associated with $Q$ and $Q_{ds}$.

We have also used a variant of the algorithm of~\cite{trevino2015} to optimize $Q_x$ in Zachary's Karate Club network~\cite{zachary1977} and the American college football network~\cite{newman2002} to detect their underlying community structure.  For the Karate Club network, Fig.~\ref{empirical}(a), many small communities are found, including four single node communities. However, each of these communities are contained within the two ground-truth communities that correspond to the historical split of the Karate Club. Thus, the community structure from maximizing $Q_x$ respects the ground truth division of the network.  In contrast, maximizing $Q_{ds}$ gives rise to a community of nodes with no internal connections that does not respect the ground truth [See Fig.~\ref{karate}(d)].

Our result for the American college football network is shown in Fig.~\ref{empirical}(b).
This is a network of games played between Division IA colleges during regular season Fall 2000~\cite{newman2002}. Community detection is usually expected to find ground-truth communities of colleges belonging to the same conference. 
As the figure shows, we find fifteen communities by optimizing $Q_x$. This contrasts with
partitions consisting of twelve communities when $Q_{ds}$ is optimized and ten communities when $Q$ is optimized. (See Supplementary Material Fig.~S1.) 
The partition that maximizes $Q_x$ detects most of the ground-truth communities, but with 
some notable exceptions. 
The partitions that maximize $Q$ and $Q_{ds}$ also show deviations from the conference membership ground-truth.
Of course, the conference structure is not necessarily the true natural structure of the network.
Some nodes, such as Texas Christian, share more links to a different conference than to the conference they are part of and are not classified correctly by conference structure. 

The partition maximizing $Q$ is farthest from the ground truth and has larger communities.
(The value of $Q$ for this partition is the same as that of the highest reported value~\cite{cafieri2011}.)
The partition maximizing $Q_{ds}$ is very similar. (This same partition was reported in~\cite{mingming2014,charo2016}.) The only differences are that the schools outside of the Southeastern Conference and the Mountain West Conference that are grouped with them now form independent communities and that Central Florida switches from grouping with the Mid-American Conference to merging with the group that split from the Mountain West Conference. Each of these differences make intuitive sense.
The partition maximizing $Q_x$ is similar to the one the maximizes $Q_{ds}$ except that the independent schools Notre Dame and Navy now split from the Big East Conference and form their own community and Connecticut splits from the Mid-American conference to form its own community. The only other, but very interesting change is that the Mid-American Conference [bottom right corner in Fig.~\ref{empirical}(b)] now splits into two communities that are eastern and the western groups. Again, each of these differences make intuitive sense. Thus, even though $Q_x$ finds smaller communities than $Q_{ds}$ or $Q$, these additional communities are meaningful. 
Furthermore, the partition that maximizes $Q_x$ in this network, as well as for the Karate Club network, is mostly consistent with those that maximize $Q_{ds}$ and $Q_x$, since the main differences simply subdivide communities found by maximizing the other metrics. 
We consider the ability of $Q_x$ to detect small groups a positive feature, not a drawback~\cite{aldecova2011}. 

\section{Conclusions}
In this paper we have discussed community detection by maximizing modularity density measures. Modularity density $Q_{ds}$ was originally introduced to address problems with modularity, most notably the resolution limit of modularity. We found that, while the use of modularity density does significantly mitigate the resolution limit problem, it does not eliminate it completely. 
In particular, we found that when using modularity density loosely connected dense communities, especially cliques, can not be resolved if they have very different sizes and constitute a large portion of the network. 
We also found that the split penalty term in modularity density can cause sets of nodes that have no common links to be grouped together as a community.  

To address these problems, we introduced a modified density metric called excess modularity density $Q_x$. We motivated the definition of the modified metric on intuitive grounds and applied it to both stylized and real-world example networks. 
We demonstrated that 
it effectively eliminates the problems associated with using both modularity and modularity density for a broad class of networks, thereby expanding the range of applicability of community detection methods.    
In the limit of a sparse network, however, excess modularity density and modularity density become equivalent and the resolution issues will also exist for $Q_x$.
Thus, despite our advances, finding a complete, general solution to the resolution limit problem remains elusive. Nevertheless, using $Q_x$ instead of $Q_{ds}$ can substantially 
improve the quality of community detection 
and we therefore propose it as a superior measure.

The metric $Q_x$ has been defined in this paper only for simple unweighted networks. Many complex networks are more complicated having, for example, weighted links and/or a bipartite structure. Definitions of modularity and modularity density have been extended to incorporate such networks by utilizing an appropriate null model~\cite{mingming2013,newman2004weighted, Barber2007}. Similar extensions can be made to excess modularity density. To use these metrics, algorithms to find the partition that maximizes them must also be developed. Such algorithms have been developed and utilized for modularity~\cite{chauhan2016, trevino2012, Bhavnani2012} and modularity density~\cite{mingming2014}. Developing such algorithms for excess modularity density would be both interesting and important.
The expected structure in the absence of any communities, i.e. the null model, plays a crucial role in determining the community structure of a given network. Usually a metric relies on a randomized network with soft constraints for this purpose. Thorough analysis of the effect of imposing hard constraints~\cite{bassler2015,orsini2015,coolen2009,charo2010,kim2012} on these null models would be another interesting topic to explore.

\section*{Acknowledgments}
This work was supported by the NSF through grants DMR-1507371 and IOS-1546858. Some of the computations in this work were done on the uHPC cluster at the University of Houston, acquired through NFS Award Number 1531814. We thank Boleslaw K. Szymanski, Charo I. Del Genio, and Weibin Zhang for fruitful discussions.

\pagebreak
\beginsupplement
\begin{center}
\section*{Supplementary Material}
\end{center}
\begin{figure}[!htb]
\centering
\subfloat[]{\includegraphics[width=.90\textwidth]{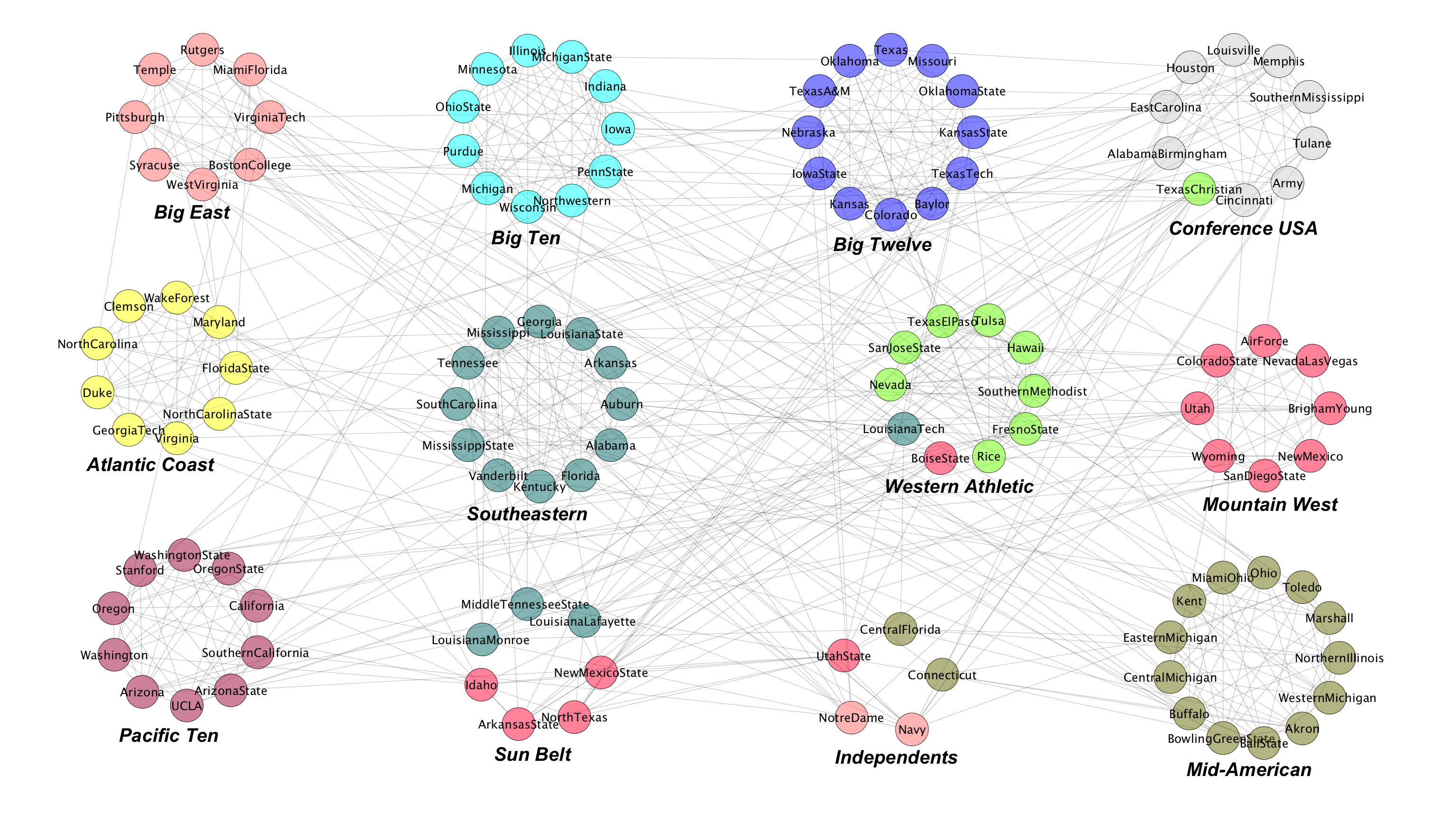}}\\
\subfloat[]{\includegraphics[width=.90\textwidth]{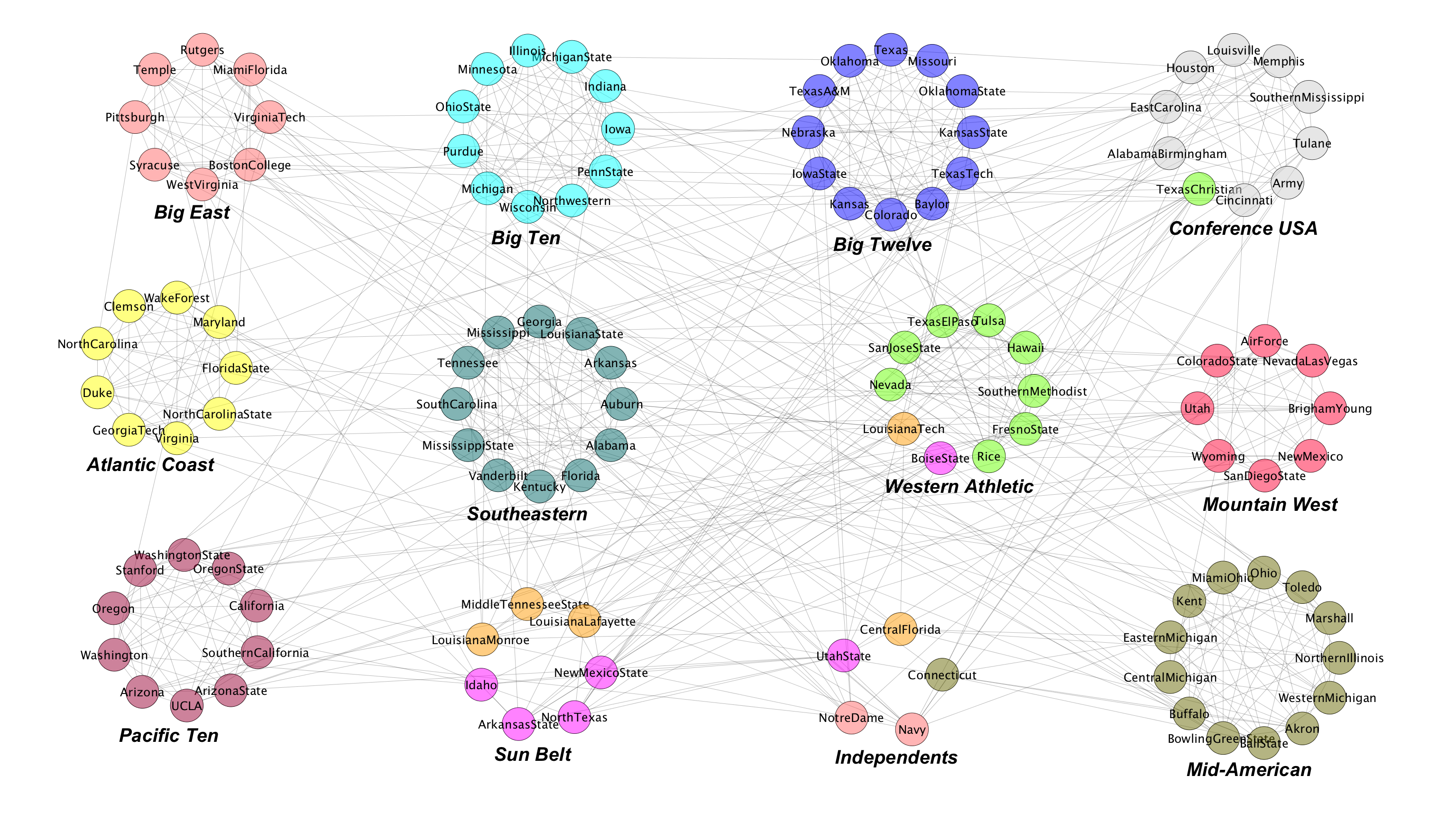}}
\caption{\label{s1} Community detection results of the American college football network. Nodes belonging to the same community are indicated by the same color. The 12 communities of the ground truth partitioning are the conferences indicated by the layout of 12 circles of nodes.  (a) Partition maximizing modularity. It has 10 communities and $Q = 0.605$. The value of $Q$ for this partition is the same as that of the highest reported value~\cite{cafieri2011}. (b) Partition maximizing modularity density. It has 12 communities and $Q_{ds} = 0.491$. This result is the same as previously reported in~\cite{mingming2014,charo2016}.}
\end{figure}
\end{document}